\documentclass[%
reprint, %two colomn option
 amsmath,
 amssymb,
 aps,
]{revtex4-2}

\usepackage{graphicx}% Include figure files
\usepackage{dcolumn}% Align table columns on decimal point
\usepackage{bm}% bold math
\usepackage{color}
\newcommand{\HgT}{HgBa$_2$Ca$_2$Cu$_3$O$_{8+ \delta}$~}
\newcommand{\BiT}{Bi$_2$Sr$_2$Ca$_2$Cu$_3$O$_{10+ \delta}$}

\begin{document}

%\preprint{APS/123-QED}

\title{Evidence for antiferromagnetism coexisting with charge order in the trilayer cuprate \HgT}

\author{V. Oliviero$^{1,\dagger}$, S. Benhabib$^{1,\dagger,\ddagger,*}$, I. Gilmutdinov$^{1}$, B. Vignolle$^{2}$, L. Drigo$^{1,\diamond}$, M. Massoudzadegan$^{1}$, M. Leroux$^{1}$, G.L.J.A. Rikken$^{1}$, A. Forget$^{3}$, D. Colson$^{3}$, D. Vignolles$^{1,*}$ and C. Proust$^{1,*}$}

\affiliation{
$^1$LNCMI-EMFL, CNRS UPR3228, Univ. Grenoble Alpes, Univ. Toulouse, INSA-T, Grenoble and Toulouse, France \\
$^2$Institut de Chimie de la Mati\`{e}re Condens\'{e}e, Bordeaux, France \\
$^3$Service de Physique de l’Etat Condensé, CEA Saclay (CNRS-URA 2464), Gif sur Yvette 91191, France\\
}

\date{\today}

\begin{abstract}
Multilayered cuprates possess not only the highest superconducting temperature transition but also offer a unique platform to study disorder-free CuO$_2$ planes \cite{Mukuda12, Kunisada20} and the interplay between competing orders with superconductivity \cite{Fradkin15, Comin15, Proust19}.  Here, we study the  underdoped trilayer cuprate \HgT \cite{Schilling93, Loret17} and we report the first quantum oscillation and Hall effect measurements in magnetic field up to 88~T. A careful analysis of the complex spectra of quantum oscillations strongly supports the coexistence of an antiferromagnetic order in the inner plane and a charge order in the outer planes. The presence of an ordered antiferromagnetic metallic state that extends deep in the superconducting phase is a key ingredient that supports magnetically mediated pairing interaction in cuprates \cite{Scalapino12}.
\end{abstract}

%\pacs{Valid PACS appear here}% PACS, the Physics and Astronomy

%\keywords{Suggested keywords}%Use showkeys class option if keyword
                              %display desired
\maketitle

%\tableofcontents

The close proximity of antiferromagnetic (AFM) order to an unconventional superconducting phase is a generic feature of strongly correlated superconductors. The coexistence and interplay of AFM order and superconductivity has led to theories based on spin-fluctuation mediated pairing interaction \cite{Scalapino12}. In cuprate high-temperature superconductors, magnetic interactions are at the heart of the debate for the pairing interaction. Although the parent compounds are antiferromagnetic Mott insulator, the presence of the pseudogap phase in hole-doped cuprates complicates the situation. Indeed, there is a variety of competing orders with superconductivity, such as charge order, stripe order or nematic phase, that nucleate inside the pseudogap \cite{Fradkin15}.
The multilayered cuprates provide a proving ground for studying such multiple phases. They have been thoroughly studied by NMR \cite{Mukuda12}, ARPES \cite{Feng02,Ideta10, Kunisada20} and Raman spectroscopy \cite{Loret19}. The highest superconducting transition temperature ($T_c$) at ambient pressure is observed for three CuO$_2$ planes \cite{Schilling93}, but the microscopic mechanism at the origin of this experimental observation is still under debate \cite{Legett99, Pavarini01, Chakravarty04, Kivelson05}. One way to understand this problem is to consider the substantial interplane coupling that could stabilize the AFM phase in the underdoped regime, thus boosting AFM fluctuations away from the ordered phase and close to optimal doping. Moreover, interplane coupling could suppress phase fluctuations and hence increase $T_c$. Another important ingredient of multilayered cuprates is the symmetry-inequivalent CuO$_2$ planes. Indeed, the fact that the inner planes (IPs) are not adjacent to the charge reservoir has two consequences: i) the inner CuO$_2$ planes are protected from out-of-plane disorder and extremely clean \cite{Mukuda12}, and ii) the fact that IPs are farther from the charge reservoir layer than outer planes (OPs) induces a charge imbalance between the different planes. This has been demonstrated by NMR measurements in several multilayered cuprates (for a review, see ref.~\onlinecite{Mukuda12}) and by ARPES measurements \cite{Ideta10} in optimally doped trilayer \BiT (Bi2223). Consequently, different competing orders can appear in the IPs and the OPs. And each of these orders could influence the Fermi surface (FS) from which high-$T_c$ superconductivity emerges at optimal doping. Namely, AFM order is known to reconstruct the FS at low doping \cite{Kunisada20}, and charge order (CO) is also now recognized as a generic property of underdoped cuprates \cite{Comin15}. For instance, in underdoped YBa$_2$Cu$_3$O$_y$ (YBCO) and HgBa$_2$CuO$_{6+\delta}$ (Hg1201), the observation of quantum oscillations (QOs) with small frequencies \cite{Doiron07, Barisic13} and negative Hall effect \cite{LeBoeuf07, Doiron13} are a strong indication of the presence of a small closed electron pocket indicating a FS reconstruction. NMR \cite{Wu11} and x-ray scattering \cite{Ghiringhelli12, Chang12, Tabis14} measurements then found evidence of CO in YBCO and Hg1201. While the exact scenario for the FS reconstruction is still debated \cite{Proust19}, a biaxial CO can indeed lead to an electron pocket in the nodal region of the first Brillouin zone \cite{Harrison12}.

%%%%%%%%%%%%%%%%%%%%%%   FIGURE 1  %%%%%%%%%%%%%%%%%%%%%
\begin{figure*} [t]
\centering
\includegraphics[width=1\textwidth]{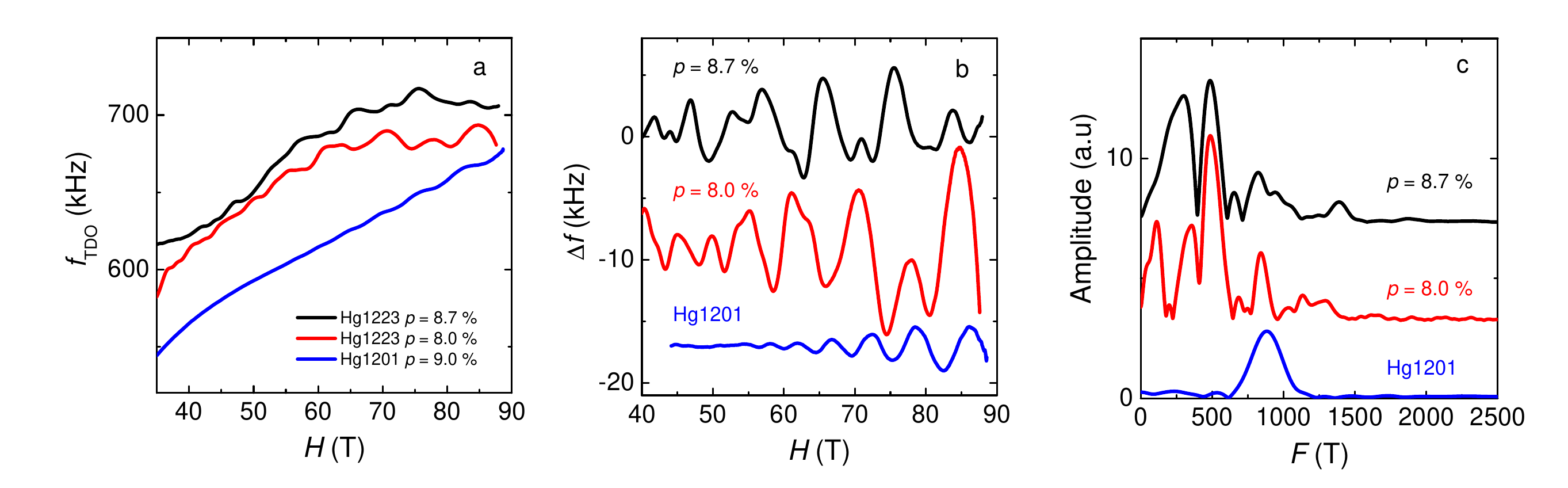}
\caption{\textbf{QO data} a) Field dependence of the TDO frequency after the heterodyne circuit at low temperature in the monolayer Hg1201 (blue line) and in the trilayer Hg1223 at different doping levels (black and red line). 
b) Oscillatory part of the TDO signal after removing a smooth background (spline) from the data shown in panel a).
c) Discrete Fourier analysis of the oscillatory part of the TDO signal shown in panel b).  
}
\label{Fig1}
\end{figure*}
%%%%%%%%%%%%%%%%%%%%%%%%%%%%%%%%%%%%%%%%%%%
Among cuprates, \HgT (Hg1223) holds the record of the highest superconducting transition temperature at ambient pressure ($T_c$=133~K). It is a trilayer cuprate and the narrow $^{63}$Cu-NMR linewidth \cite{Julien96, Mukuda12} clearly shows that the IP is extremely clean as it is homogeneously doped and screened from out-of-plane disorder by the OPs. In addition, Raman spectroscopy shows the typical signature of CO in optimally and underdoped Hg1223 \cite{Loret19}, and NMR measurements performed in an equivalent trilayer cuprate \cite{Mukuda12} suggest the critical doping at which AFM order in the IP disappears corresponds to an average carrier density $p$ = 9~\% ($T_c \approx$ 80~K). 
Here, we investigate the transport properties of underdoped Hg1223 in the doping range $p$ = 8 - 8.8~\% by means of contactless resistance and Hall effect measurements in pulsed fields up to 88~T. We discover quantum oscillations with small frequencies and a Hall coefficient that remains positive down to the lowest temperature, evidencing the presence of small reconstructed pockets of both holes and electrons, which strongly supports the coexistence of AFM and CO but on different CuO$_2$ planes, with AFM on the IPs and CO on the OPs. An additional frequency corresponding to magnetic breakdown tunnelling between the inner and outer planes is also observed.\\
Fig.~\ref{Fig1}a shows the variation of the tunnel diode oscillator (TDO) circuit frequency (see Methods) as a function of magnetic field for Hg1201 at $p$ = 9\% and for two samples of Hg1223 at slightly different doping levels. In the latter, QOs are clearly observed above $H$ = 40~T, confirming the high quality of the samples. A smooth background subtraction leads to the oscillatory part of the signal shown in Fig.~\ref{Fig1}b. While there is obviously only one QO frequency for Hg1201, the QO spectrum of Hg1223 is much more complex and contains several frequencies. This is confirmed by the discrete Fourier transform  analysis depicted in Fig.~\ref{Fig1}c. In Hg1201, the discrete Fourier transform reveals a single frequency $F$ = 850~T, in agreement with previous studies \cite{Barisic13, Chan16}. For Hg1223, neglecting the low frequencies that can be attributed to imperfect background subtraction, at least three frequencies can be isolated at $F_1 \approx $~350~T, $F_2 \approx $~500~T and $F_3 \approx $~850~T, where $F_3 - F_2 \approx F_1$. Some harmonics and frequency combinations are also present at higher frequencies. The temperature dependences of the QOs are shown in the SI, Fig.~\ref{FigS2}. As expected from the Lifshitz-Kosevich theory \cite{Shoenberg}, the amplitude of QOs decreases as the temperature increases and vanishes above $T \approx $ 10~K. 
%
%%%%%%%%%%%%%%%%%%%%%%   FIGURE 2  %%%%%%%%%%%%%%%%%%%%%
\begin{figure}
\centering
\includegraphics[width=0.45\textwidth]{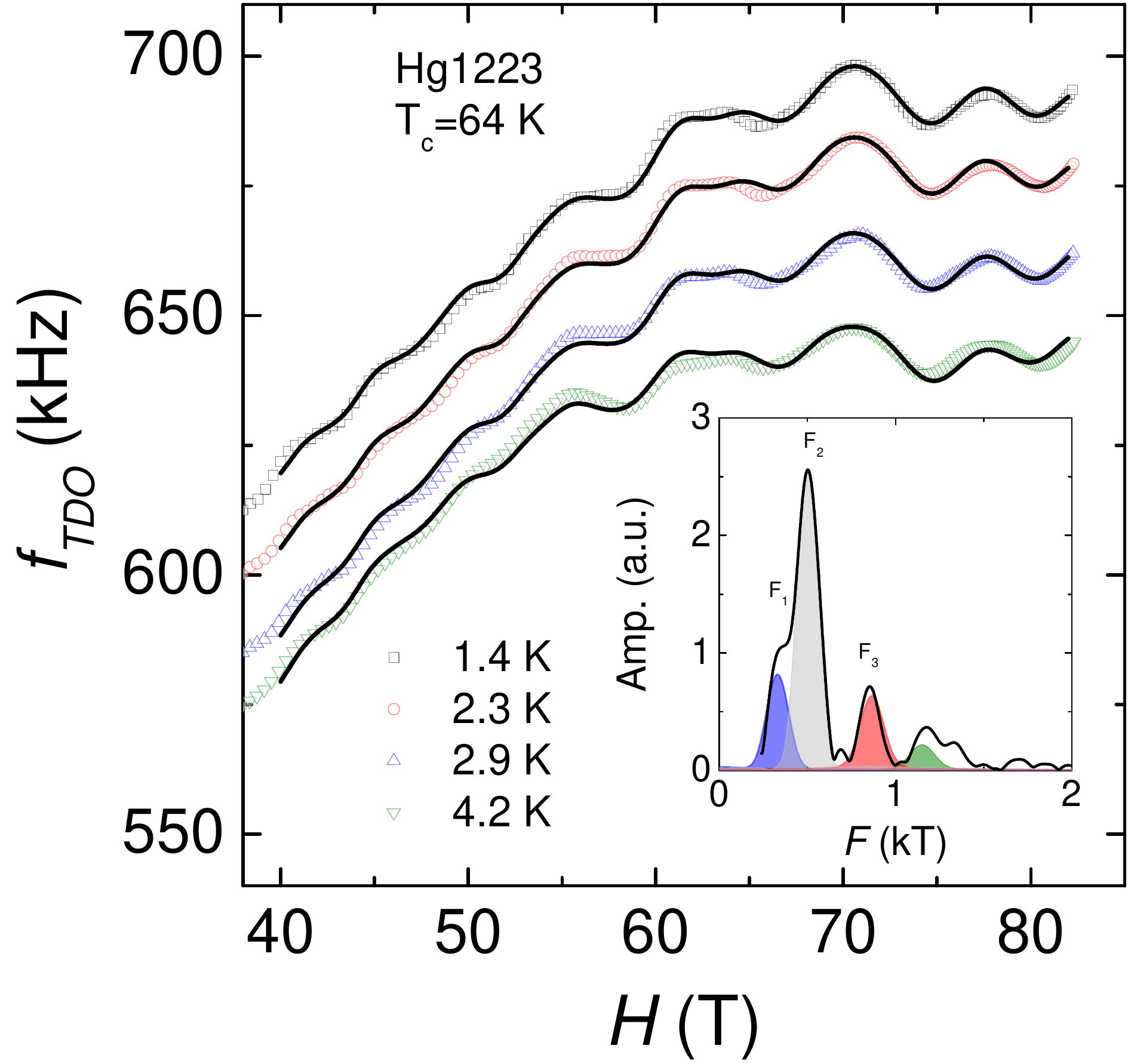}
\caption{\textbf{Lifshitz-Kosevich fits} Field dependence of the TDO frequency in Hg1223 ($p$ = 8~\%) at different temperatures (symbols). Solid lines correspond to the fits to the data using the Lifshitz-Kosevich theory plus a polynomial background in the field range $40 \leq H \leq 83$~T (see SI for details).
The inset shows the Fourier analysis of the oscillatory part of the data at $T$ = 1.5~K along with the contribution of $F_{1}$ (blue), $F_2$ (grey) and $F_3$ (red), respectively. The green component corresponds to a frequency combination, which has been taken into account to improve the fits. 
}
\label{Fig2}
\end{figure}
%%%%%%%%%%%%%%%%%%%%%%%%%%%%%%%%%%%%%%%%%%%
%%%%%%%%%%%%%%%%%%%%%%   TABLE 1  %%%%%%%%%%%%%%%%%%%%%
\begin{table}[h]
\centering
%\begin{tabular*}{1\textwidth}{|c|c|c|c|c|c|}
 \begin{tabular}{|c|c|c|c|c|c|}
  \hline
  Family & $T_c$  & $p$ & $F_1$ & $F_2$ & $F_3$ \\
  \hline
  Hg1223 & 64~K & 8.0~\% & 330$\pm$30~T & 500$\pm$20~T & 850$\pm$20~T \\
  \hline
  Hg1223 & 74~K & 8.7~\% & 335$\pm$20~T & 500$\pm$20~T & 850$\pm$20~T \\
  \hline
  Hg1201 & 74~K & 9.7~\% & \multicolumn{2}{|l|}{} & 880 T  \\
  \hline
\end{tabular}
\caption{\label{tab1} Sample family, $T_c$, doping $p$ and QO frequency deduced from the discrete Fourier transform analysis.  }
\end{table}

A challenge in analysing these data is that the oscillation frequencies are low and there is a limited field range available.
Therefore, the accurate determination of the value of the frequencies is ambiguous, in particular for the nearby frequencies $F_1$ and $F_2$. To assert the spectra of QOs, we performed fits to the data at different temperatures using the Lifshitz-Kosevich theory (see SI for the detailed procedure of the fit). In order to constrain the fits, we performed simultaneous fits to the dataset at different temperatures, where all parameters (except for the background) are temperature independent. Fig.~\ref{Fig2} shows the raw data at different temperatures for the sample at $p$ = 8~\% (symbols) and solid lines are the simultaneous fits (see Supplementary Fig.~\ref{FigS3} for the $p$ = 8.7~\% sample). The value of the frequencies deduced from the fitting procedure at $p$ = 8~\% are $F_1$ = 331~T, $F_2$ = 500~T and $F_3$ = 866~T, in good agreement with the values obtained by discrete Fourier transform at different temperatures (see Table~\ref{tab1}). Both analysis confirmed that the oscillatory spectrum is composed of at least three frequencies linked by the relation $F_3 - F_2 \approx F_1$. 

%%%%%%%%%%%%%%%%%%%%%%   FIGURE 3  %%%%%%%%%%%%%%%%%%%%%
\begin{figure} %[h]
\centering
\includegraphics[width=0.45\textwidth]{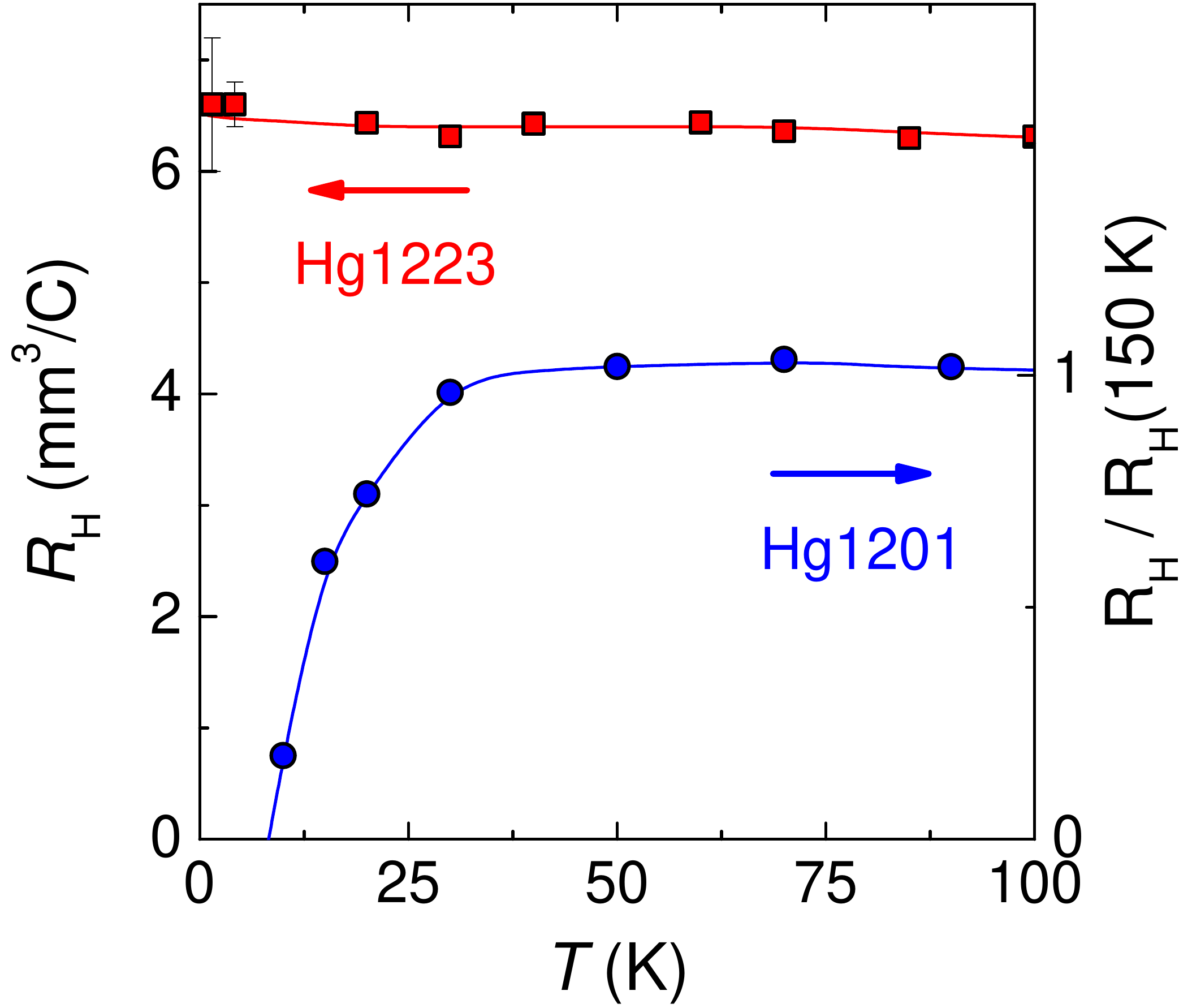}
\caption{\textbf{Hall data} Temperature dependence of the normal-state Hall coefficient $R_H$, measured at high fields, in Hg1223 ($p$ = 8.8~\%, red squares) and in Hg1201 ($p$ = 8~\%, blue circles adapted from ref.~\onlinecite{Doiron13}). The Hall coefficient changes sign in Hg1201 while it remains positive (i.e. hole-like) down to the lowest temperature in Hg1223. Note that in Hg1223, as $T \rightarrow$ 0, $R_H \approx$ 6.5~mm$^3$/C, corresponding to an effective carrier density $p_H \approx$ 8~\%.
}
\label{Fig3}
\end{figure}
%%%%%%%%%%%%%%%%%%%%%%%%%%%%%%%%%%%%%%%%%%%
In order to gain more insight in the Fermi surface of underdoped Hg1223, we performed Hall effect measurements up to 88~T at a doping level $p$ = 8.8~\%. Fig.~\ref{Fig3} shows the temperature dependence of the normal-state Hall coefficient down to $T$ = 1.5 K (the isotherms are shown in the SI, Fig.~\ref{FigS4}). Remarkably, there is almost no temperature dependence of the Hall coefficient and it remains positive down to the lowest temperatures. This result contrasts with the Hall coefficient in underdoped YBCO \cite{LeBoeuf07, Badoux16} and Hg1201 \cite{Doiron13}, that changes sign and becomes negative at low temperatures. This has been interpreted as the signature of an electron pocket resulting from the FS reconstruction caused by the CO. In the case of underdoped YBCO, the CO is present in both CuO$_2$ planes of the bilayer and QO measurements reveal the main frequency $F_e$ = 540~T flanked by two nearby and equally spaced satellites $F_e \pm $ 90~T at a doping level $p \approx$ 11~\% \cite{Audouard09, Sebastian12}. This has been interpreted as magnetic breakdown tunnelling between bilayer-split pockets \cite{Sebastian12, Maharaj16, Briffa16} provided that the mirror symmetry between the planes of the bilayer is broken. In the single layer Hg1201, only one QO frequency $F \approx$ 850~T has been detected so far \cite{Barisic13, Chan16}. The difference in the value of the main frequency between YBCO and Hg1201 is a direct consequence of a different CO wavevector, leading to different size of the reconstructed electron pocket.

We now discuss different scenarios to explain our results in underdoped Hg1223. 

(i) \emph{Band structure calculation}

Local-density-approximation calculations \cite{Singh93} of the electronic structure of the stoichiometric compound HgBa$_2$Ca$_2$Cu$_3$O$_8$ reveal that the Fermi surface consists of three large hole-like tubular CuO$_2$ sheets centered on the corner of the Brillouin zone plus a small electron-like Fermi surface located at the anti-node (see Fig. 10a). The latter disappears with doping \cite{Singh93}. For a doping $p \approx$ 8~\%, the Fermi surface of the CuO$_2$ sheets corresponds to $1 + p$ carriers that translates into a QO frequency $F_{LDA} \approx$ 15~kT much larger than the observed frequencies in our study.

(ii)  \emph{Charge order in the three CuO$_2$ planes}

In analogy with underdoped YBCO where a CO is present in both CuO$_2$ planes, let us assume that a CO is present in the three CuO$_2$ planes of Hg1223 (see a sketch of the scenario in the SI, Fig. 10b). As the frequencies are not equally spaced, the model of magnetic breakdown tunnelling is inadequate to explain the spectrum of oscillation frequencies in Hg1223. Moreover, the Hall coefficient remains positive down to the lowest temperature, in contrast with underdoped Hg1201 and YBCO.

(iii) \emph{AFM in the inner plane}

Another scenario assumes an AFM metallic phase in the IP and a corresponding FS that contains both electron and hole pockets (see a sketch of this scenario in the SI, Fig. 10c). While there is no direct evidence yet of an AFM order in underdoped Hg1223, such order has been detected by extensive NMR measurements in the IP of the trilayer cuprate Ba$_2$Ca$_2$Cu$_3$O$_6$(F,O)$_2$ (0223F) with $T_c$ up to 81~K but not beyond. \cite{Shimizu11, Mukuda12} Given the disorder-protected nature of IPs in multilayered cuprates, let us assume that QOs originate from quasi-particles in the IP with $F_{hole}$=$F_3 \approx$ 850~T and $F_{electron}$=$F_2 \approx$ 500~T. In this scenario, the third frequency $F_1$=$F_3$-$F_2$ would correspond to magnetic breakdown between the hole and the electron pockets in the IP. However, in order to reproduce the size of the orbits corresponding to the observed frequencies $F_2$ and $F_3$, the AFM potential used in the calculation is 0.45 eV (see discussion in the SI and Fig.~\ref{FigS6}c). This value translates to a magnetic breakdown field unattainable, ruling out the possibility to observe magnetic breakdown between the hole and electron pocket. Finally, the presence of an electron pocket at the anti-node is difficult to reconcile with the presence of a pseudogap.
%
%%%%%%%%%%%%%%%%%%%%%%   FIGURE 4  %%%%%%%%%%%%%%%%%%%%%
\begin{figure} %[h]
\centering
\includegraphics[width=0.52\textwidth]{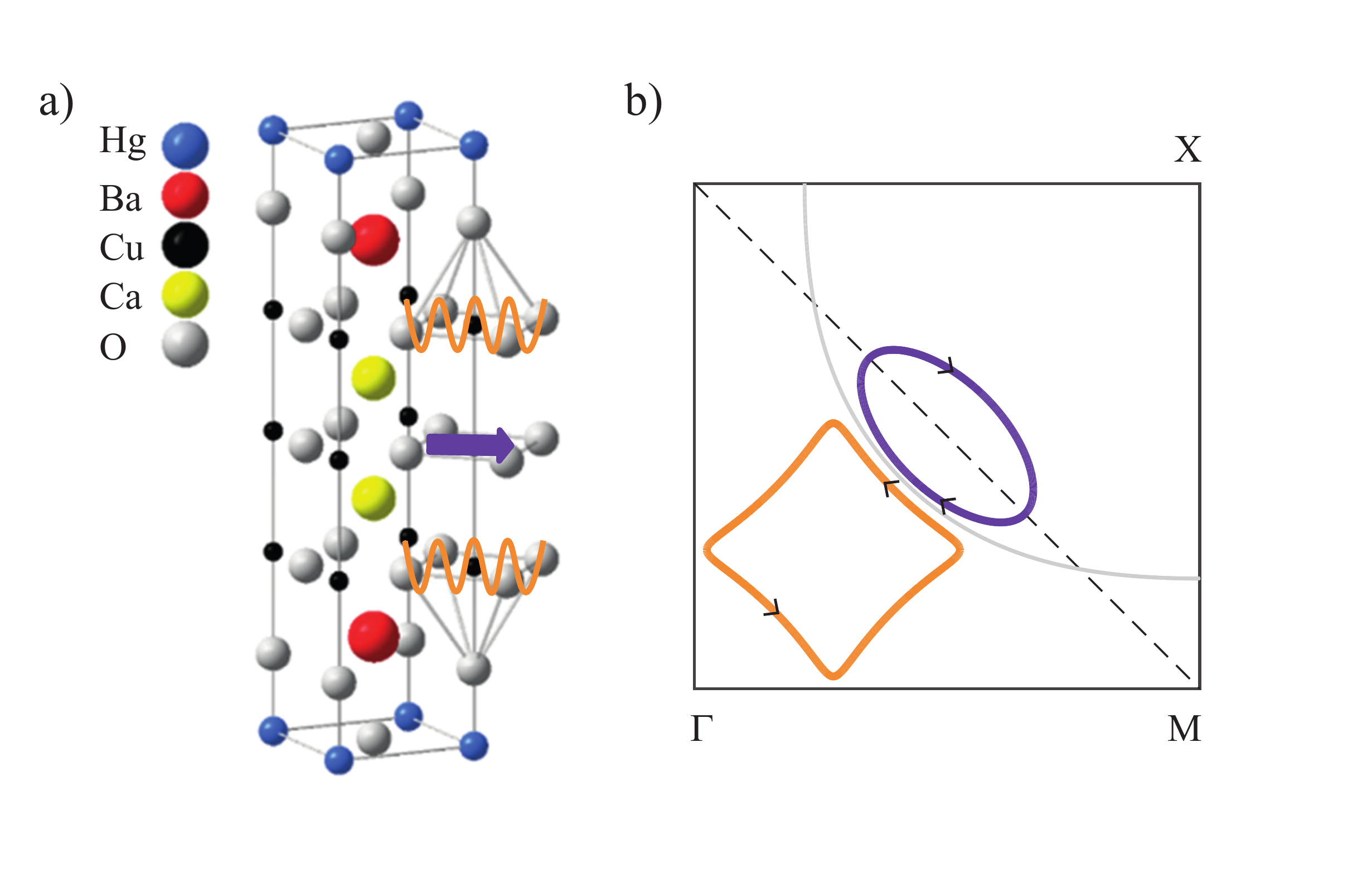}
\caption{\textbf{Sketch of the Fermi surface} a) Crystallographic structure of trilayer Hg1223. We sketch the presence of AFM order in the IP (purple arrow) and charge order (orange wave) in the OPs. b) Corresponding reconstructed Fermi surface in presence of AFM order in the IP leading to a hole pocket (purple, $F_2$ QO frequency) and CO order in the OP leading to an electron pocket (orange, $F_3$ QO frequency). Both pockets are located in the nodal region of the quarter of the first Brillouin zone. Magnetic breakdown tunnelling between the pockets leads to an additional frequency $F_1 \approx F_3 - F_2$.
}
\label{Fig4}
\end{figure}
%%%%%%%%%%%%%%%%%%%%%%%%%%%%%%%%%%%%%%%%%%%

(iv) \emph{AFM in the inner plane and charge order in the outer plane}

Given the charge imbalance between IP and OP, the carrier density is always lower in the IP. Let us assume that an AFM metallic phase is present in the IP in analogy with the 0223F compound. \cite{Shimizu11, Mukuda12} But compared to scenario (iii), the FS in the IP consists solely of hole pockets at the nodes corresponding to $F_{hole}$=$F_2 \approx$~500~T. This is in agreement with recent ARPES and QOs studies showing the metallic character of the AFM phase at low doping in the IPs of a 5-layer cuprate\cite{Kunisada20}. Kunisada \textit{et al.} found two QO frequencies $F$(IP0) = 147~T and $F$(IP1) = 318~T corresponding to an effective carrier density $p$ = 2.1~\% and $p$ = 4.5~\%, respectively. In our study, the hole frequency $F_2 \approx$~500~T translates to a carrier density $p$ = 7.2~\%, in good agreement with the estimation given by NMR measurements in a 3-layer cuprate with $T_c$ = 76~K, where $p$(IP) $\approx$ 7.4~\% and $p$(OP) $\approx$ 8.7~\% \cite{Mukuda12}. In addition, as shown by recent Raman spectroscopy measurements in underdoped Hg1223 \cite{Loret19}, we assume that the CO sets in the OP as the effective doping is higher. It induces a Fermi surface reconstruction leading to an electron pocket at the node corresponding to $F_{electron}$=$F_3 \approx$~850~T, in analogy with the monolayer Hg1201. In Fig.~\ref{Fig4}a, we sketch the real space structure corresponding to this scenario: an AFM order in the IP coexisting with a charge order in the OPs. Fig.~\ref{Fig4}b shows the resulting Fermi surface consisting of both electron (orange) and hole (purple) pockets. The third frequency $F_{MB}$=$F_e$-$F_h$ would correspond to magnetic breakdown tunnelling between OP and IP. But how can we reconcile this scenario with a positive Hall effect? Let us focus on the low temperature value of the Hall coefficient and assume the low-field limit for the two-band model of the Hall effect (see in the SI for a discussion). The Hall coefficient is given by: 
$R_H =\frac{\sigma_h \mu_h - \sigma_e \mu_e}{(\sigma_h + \sigma_e)^2}$, where $\sigma$ and $\mu$ are the conductivities and mobilities, respectively. Given the carrier densities deduced from quantum oscillation frequencies, a Hall coefficient $R_H \approx $ 6.5 mm$^3$/C (see Fig.~\ref{Fig3}) corresponds to a ratio of mobilities $\mu_h / \mu_e \approx $ 3, a reasonable value owing to the disorder-protected nature of the IP compared to the OPs. In short, this interpretation allows to explain both the QO spectrum and the value of $R_H$ at low temperature. \\
Our interpretation implies that, in the cuprate where $T_c$ is maximum among all superconductors, a metallic AFM state extends deep inside the SC phase. This is reminiscent of a quantum critical point scenario observed in other unconventional superconductors, where spin fluctuations extend away from the AFM ordered state. The dispersion of such magnetic excitations have first been measured in YBCO using inelastic neutron scattering \cite{Hayden04}. Resonant inelastic x-ray scattering (RIXS) experiments have subsequently extracted the dispersion of these magnetic excitations, called paramagnons, up to high energy transfer in different cuprate families and over a large doping range \cite{LeTacon11}. Interestingly, a recent RIXS study on the two first member of the Hg-family of cuprates shows that the energy scale of the paramagnon spectra matches with the ratio of $T_c$ \cite{Wang20}. All of the above considerations strongly suggest a magnetic pairing mechanism for cuprates. In Hg1223, the clean nature and the absence of buckling of the inner CuO$_2$ plane support the idea that the antiferromagnetic interaction $J$ is large, leading to higher $T_c$ \cite{Julien96, Wang20}. 
Could the presence of charge order in the OPs be a consequence of charge imbalance and / or of out-of-plane disorder? In YBCO, charge order competes both with SC and AFM order. This could explain why $T_c$ further increases in optimally doped Hg1223 by applying pressure \cite{Chu93} which destabilizes charge order \cite{Cyr-Choiniere18}.

%%%%%%%%%%%%%%%%%  REFERENCES  %%%%%%%%%%%%%%%%%%%%%%%%%%%%%%%%%%
%

\clearpage
\section*{Methods}
\subsection*{Samples}
Single crystals of the trilayer cuprate \HgT have been synthesized using a self-flux growth technique as described in ref.~\onlinecite{Loret17}. Using adequate heat treatment, Hg1223 can be largely underdoped and its doping level controlled.  The doping $p$ has been deduced from the empirical relation $1 - T_c / T_{c,max} = 82.6(p - 0.16)^2$, where $T_c$ is the onset superconducting transition measured by SQUID (see Fig. 5) and $T_{c,max}$ = 133~K.\\
\subsection*{TDO measurements}
Quantum oscillations have been measured using a contactless tunnel diode oscillator-based technique \cite{Coffey00}. The experimental setup consists of a LC-tank circuit powered by a tunnelling diode oscillator biased in the negative resistance region of the current-voltage characteristic. The sample is placed in a compensated 8-shape coil (diameter and length of the coil are adapted for each sample to optimize the filling factor). The fundamental resonant frequency $f_0$ of the whole circuit is around 25 MHz. The RF signal is amplified and demodulated down to a frequency of about 1~MHz using a heterodyne circuit. A high-speed acquisition system is used to digitize the signal. The data are post-analysed using a software to extract the field dependence of the resonance frequency $f_{TDO}$, which is sensitive to the resistivity through the change in skin depth. 
\subsection*{Hall effect measurements}
For Hall effect measurements, gold contacts were sputtered onto the surface of the sample before a heat treatment leading to contact resistances of a few ohms. The magnetic field $H$ was applied along the $c$-axis of the tetragonal structure, perpendicular to the CuO$_2$ planes in both polarity of the field.
The high temperature measurements were performed in a conventional pulsed magnet up to 68~T down to 10~K. At lower temperature, higher magnetic fields were required to quench superconductivity, so we extended our measurements up to 88 T, using a dual coil magnet.
The pulsed-field measurements were performed using a conventional 4-point configuration with a current excitation of 5~ mA at a frequency of ~ 60 kHz. A high-speed acquisition system was used to digitize the reference signal (current) and the voltage drop across the sample at a frequency of 500 kHz. The data was post-analyzed with a software to perform the phase comparison.\\
%\textbf{Acknowledgements}\\
%We thank  for helpful and stimulating discussions.

%C.P. acknowledges support from the EUR grant NanoX n\textsuperscript{o}ANR-17-EURE-0009 and from the ANR grant NEPTUN n\textsuperscript{o}ANR-19-CE30-0019-01. 
%\\
\textbf{Author Contributions}
SB and LD upgraded the TDO setup up to 88 T with the contribution from CP.
SB performed the TDO measurements in Hg1223 with the help of DV and CP.
DV performed the TDO experiments in Hg1201 with the help of CP.
SB performed the preliminary TDO data analysis. VO did the detailed TDO data analysis and the fits shown in the manuscript, with the help of CP.
IG and VO performed the Hall effect measurements with the help of DV and CP. IG did the analysis of the Hall effect data.
BV performed the Fermi surface reconstruction calculations.
AF and DC grew, annealed the single crystals and performed SQUID measurements.
All authors provided critical feedback and helped shape the research and analysis. CP supervised the project and wrote the manuscript with inputs from all the authors. \\
\\
\textbf{Competing interests}
The authors declare no competing interests.\\
\\
\textbf{Data availability}
All data that support the findings of this study are available from the corresponding authors on request.\\
\\
$\dagger$ These authors contributed equally to this work.\\
$\ddagger$ Present address: Institute of Physics, EPFL, CH-1015 Lausanne, Switzerland.\\
$\diamond$ Present address: GET (UMR5563 CNRS, IRD, Univ. Paul Sabatier, CNES), 31400 Toulouse, France.\\
\newline
$*$ Correspondence and requests for materials should be addressed to S.B. (siham.benhabib@epfl.ch), D.V. (david.vignolles@lncmi.cnrs.fr) or C.P. (cyril.proust@lncmi.cnrs.fr).

%%%%%%%%%%%%%%%%%%%%%%%%%%%%%%%%%%%%%%%%%%%%%%%%%%%%%%%%%%%%%%%%%%%
%%%%%%%%%%%%%%%%%%%SUPPLEMENTARY information %%%%%%%%%%%%%%%%%%%%%%
%%%%%%%%%%%%%%%%%%%%%%%%%%%%%%%%%%%%%%%%%%%%%%%%%%%%%%%%%%%%%%%%%%%

% ****** End of file ******
% ****** Start of file apssamp.tex ******
%
%   This file is part of the APS files in the REVTeX 4.1 distribution.
%   Version 4.1r of REVTeX, August 2010
%
%   Copyright (c) 2009, 2010 The American Physical Society.
%
%   See the REVTeX 4 README file for restrictions and more information.
%
% TeX'ing this file requires that you have AMS-LaTeX 2.0 installed
% as well as the rest of the prerequisites for REVTeX 4.1
%
% See the REVTeX 4 README file
% It also requires running BibTeX. The commands are as follows:
%
%  1)  latex apssamp.tex
%  2)  bibtex apssamp
%  3)  latex apssamp.tex
%  4)  latex apssamp.tex
%

%\usepackage[showframe,%Uncomment any one of the following lines to test
%%scale=0.7, marginratio={1:1, 2:3}, ignoreall,% default settings
%%text={7in,10in},centering,
%%margin=1.5in,
%%total={6.5in,8.75in}, top=1.2in, left=0.9in, includefoot,
%%height=10in,a5paper,hmargin={3cm,0.8in},
%]{geometry}

\newpage

%\preprint{APS/123-QED}

\section*{Supplementary Information }

%\begin{abstract}
%\end{abstract}

%\begin{widetext}

\section{Magnetization measurements}
\begin{figure} [h] 
\centering
\includegraphics[width=0.45\textwidth]{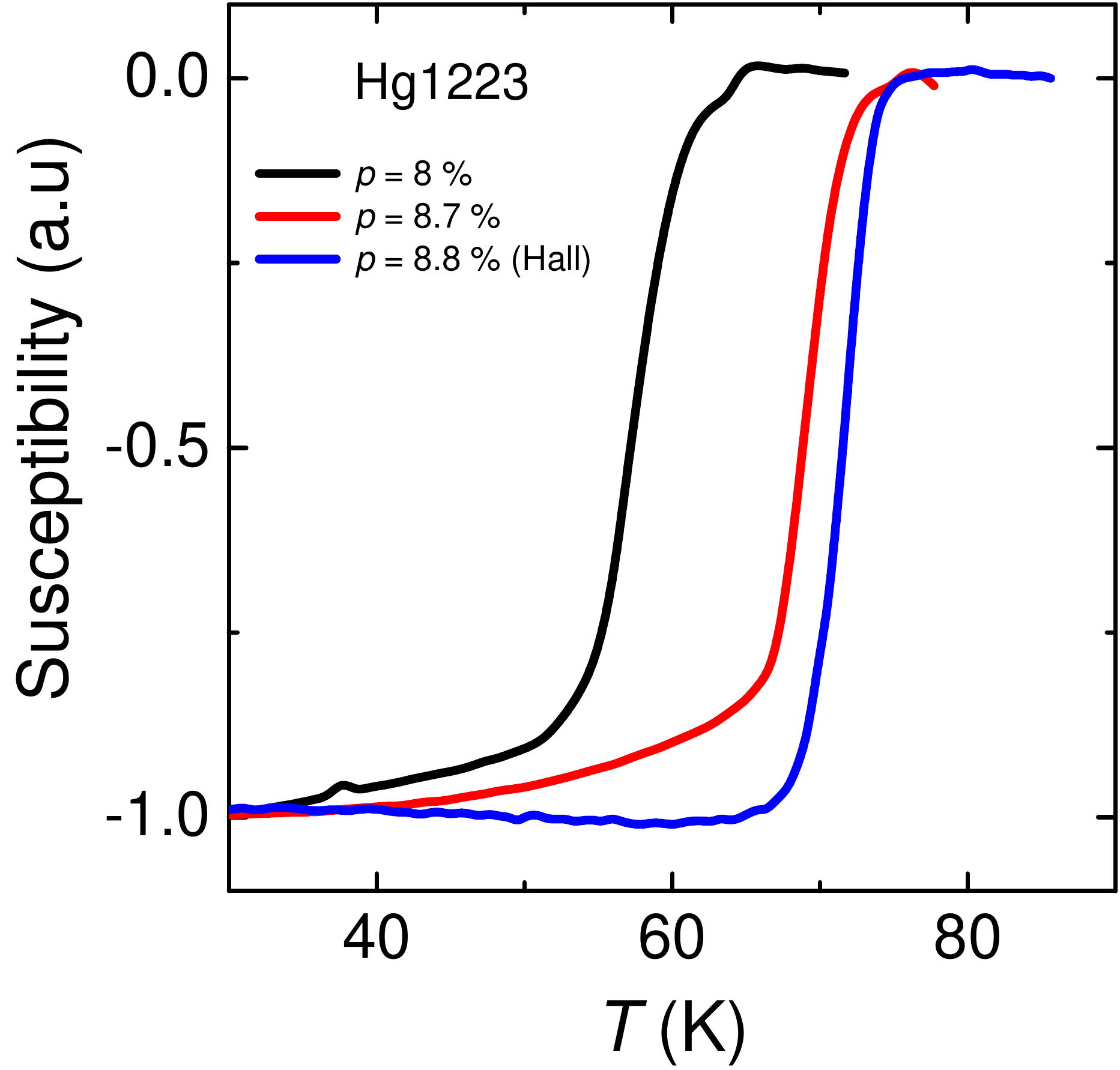}
\caption{\textbf{SQUID data.} Magnetic susceptibility measurements of vacuum-annealed crystals of \HgT at different doping levels, as indicated. 
}
\label{FigS1}
\end{figure}

Fig.~\ref{FigS1} shows the temperature dependence of the magnetic susceptibility of the three samples of \HgT. $T_c$ corresponds to the onset temperature where the susceptibility starts to drop. The $p$ = 8~\% (black line) and $p$ = 8.7~\% (red line) have been studied by quantum oscillation measurements (TDO). Hall effect measurements have been performed on the $p$ = 8.8~\% sample (blue line). 
%%%%%%%%%%%%%%%%%%%%%%   FIGURE S1  %%%%%%%%%%%%%%%%%%%%%
%%%%%%%%%%%%%%%%%%%%%%%%%%%%%%%%%%%%%%%%%%%

\section{Quantum oscillation data}
\begin{figure*} [] 
\centering
\includegraphics[width=0.9\textwidth]{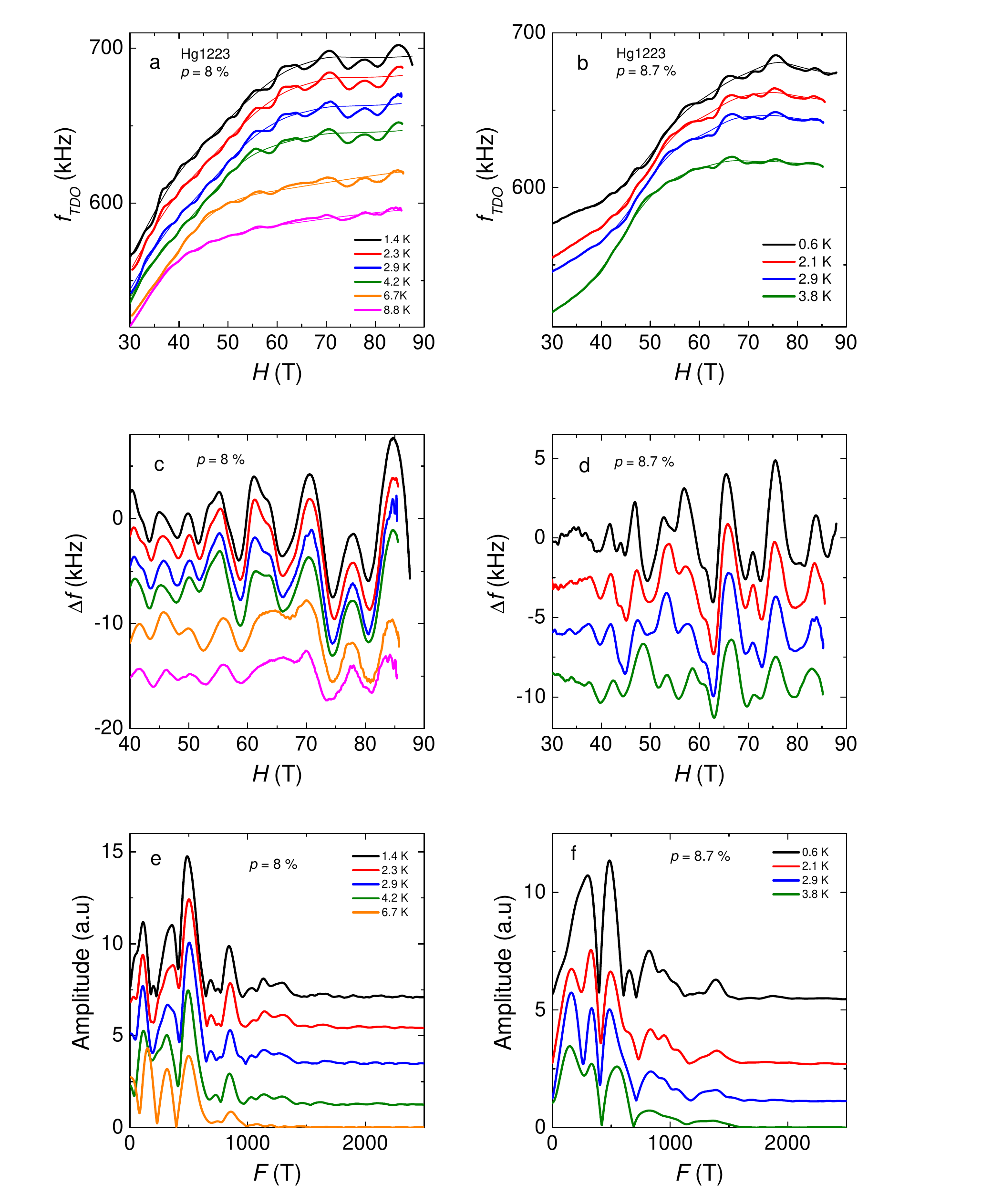}
\caption{\textbf{TDO data.} {a) Field dependence of the TDO frequency after the heterodyne circuit at low temperature in the trilayer Hg1223 $p$ = 8~\% at different temperatures, as indicated. The thin lines correspond to the background (spline) of the TDO signal. b) Same graph but for our Hg1223 $p$ = 8.7~\% sample.  
Oscillatory part of the TDO signal after removing a smooth background (spline) for c) $p$ = 8~\% and d) $p$ = 8.7~\% .
e) Discrete Fourier analysis in the field range $H$ = 35 - 85.5~T of the data shown in panel c) for $p$ = 8~\%. 
f) Discrete Fourier analysis in the field range $H$ = 37 - 85.3~T of the data shown in panel d) for $p$ = 8.7~\%.
}}
\label{FigS2}
\end{figure*}

The complete data of quantum oscillations in two samples of Hg1223 is displayed in Fig.~\ref{FigS2}. Fig.~\ref{FigS2}a and Fig.~\ref{FigS2}b show the variation of the Tunnel Diode Oscillator (TDO) frequency as a function of magnetic field up to $H$ = 88~T at different temperatures for the samples $p$ = 8~\% and $p$ = 8.7~\%, respectively. The thin lines correspond to the smooth background (splines) subtracted from the raw data leaving the oscillatory part of the signal depicted in Fig.~\ref{FigS2}c ($p$ = 8~\%) and Fig.~\ref{FigS2}d ($p$ = 8.7~\%). Figs.~\ref{FigS2}e and Fig.~\ref{FigS2}f show the discrete Fourier transform (DFT) of the oscillatory part in the field range [35~T, 85.5 T] ($p$ = 8~\%) and [37~T, 85.3 T] ($p$ = 8.7~\%), respectively.
%
%%%%%%%%%%%%%%%%%%%%%%   TABLE 1  %%%%%%%%%%%%%%%%%%%%%
\begin{table}[h]
\centering
%\begin{tabular*}{1\textwidth}{|c|c|c|c|c|c|}
 \begin{tabular}{|c|c|c|c|c|c|}
  \hline
  Family & $T_c$  & $p$ & $m_{c1}^*$ & $m_{c2}^*$ & $m_{c3}^*$ \\
  \hline
  Hg1223 & 64~K & 8.0~\% & 0.75$\pm$0.2 & 1.0$\pm$0.2 & 1.5$\pm$0.2 \\
  \hline
  Hg1223 & 74~K & 8.7~\% & N/A & 2.2$\pm$0.3 &  2.4$\pm$0.2 \\
  \hline
\end{tabular}
\caption{\label{mc} Effective masses deduced from the DFT analysis for the Hg1223 samples.  }
\end{table}  
%
%%%%%%%%%%%%%%%%%%%%%%   FIGURE S2  %%%%%%%%%%%%%%%%%%%%%
%%%%%%%%%%%%%%%%%%%%%%%%%%%%%%%%%%%%%%%%%%%

\section{Fitting quantum oscillations}
We fit the field dependence of the TDO signal to $f_{TDO} = (a_0 + a_1 H + a_2 H^2+...)+ \Delta f_{osc}$, where the first term is a polynomial representing the non-oscillating background and $\Delta f_{osc}$ is given by the Lifshitz-Kosevich (LK) theory,
\begin{align}
   \Delta f_{osc} = \sum_{i}{A_{0i} R_{Ti} R_{Di} sin[2\pi(\frac{F_i}{H}-\gamma_i)]}
    \label{eq:1}
\end{align}
where $A_{0i}$ are prefactors, $F_i$ are the oscillation frequencies and $\gamma_i$ are the phase
factors. We neglect any contribution from magnetic breakdown and spin damping.
%or any additional damping coming from the effect of superconductivity on the amplitude of QOs
$R_{Ti}$ and $R_{Di}$ are the thermal ($R_{Ti} = \alpha T
m_i^*/Hsinh[\alpha T m_i^{*}/H]$) and Dingle ($R_{Di} = exp[-\alpha T_{Di}
m_i{^*}/H]$) damping factors, respectively, where $\alpha =
2\pi^2k_Bm_0/e\hbar$ ($\simeq$ 14.69~T/K), $m_i^*$ and $T_{Di}$ are the cyclotron masses and the Dingle
temperatures, respectively \cite{Shoenberg}.
Some of the scattering going into $T_D$ probably comes from scattering due to vortices, but we leave this as a field-independent scattering contribution for simplicity.            
For each frequency, there are five free parameters. In order to constrain the parameters, we perform simultaneous fits with equation (1) to the total data set at different temperatures, where all parameters are temperature independent. 
We use a gradient based search algorithm where the optimal solution is the set of parameters that gives the smallest least squares value. Due to local minima, there exist a large number of solutions and we have constrained a few parameters like the effective masses (to agree with the DFT analysis) and the Dingle temperature (to take into account the disorder protected nature of the inner plane).  
Fig.~2 shows the raw data (symbols) for the $p$ = 8~\% sample along with the results of the fitting procedure (black lines) at different temperatures from $T$=1.4~K to $T$=4.2~K and in the field range $H$ = 40 - 83~T. Table~\ref{tabS3} shows the relevant parameters deduced from the fit, e.g. frequencies, effective masses and Dingle temperatures. The contribution of each frequency to the total signal is seen in the DFT depicted in the inset of Fig.~2. In addition to the fundamental frequencies, $F_1$ = 331~T (blue), $F_2$ = 500~T (grey) and $F_3$ = 866~T (red), a harmonic of the signal at $F_4$ = 1150~T has been taken into account to improve the fit. The simulated TDO signal does not reproduce  perfectly the data due to a complicated background and additional harmonics of the signal. Nevertheless, the oscillation frequencies  match the value obtained by the DFT within the error bars (see table~I of the main text).

%%%%%%%%%%%%%%%%%%%%%%   FIGURE S3  %%%%%%%%%%%%%%%%%%%%%
\begin{figure} [] 
\centering
\includegraphics[width=0.45\textwidth]{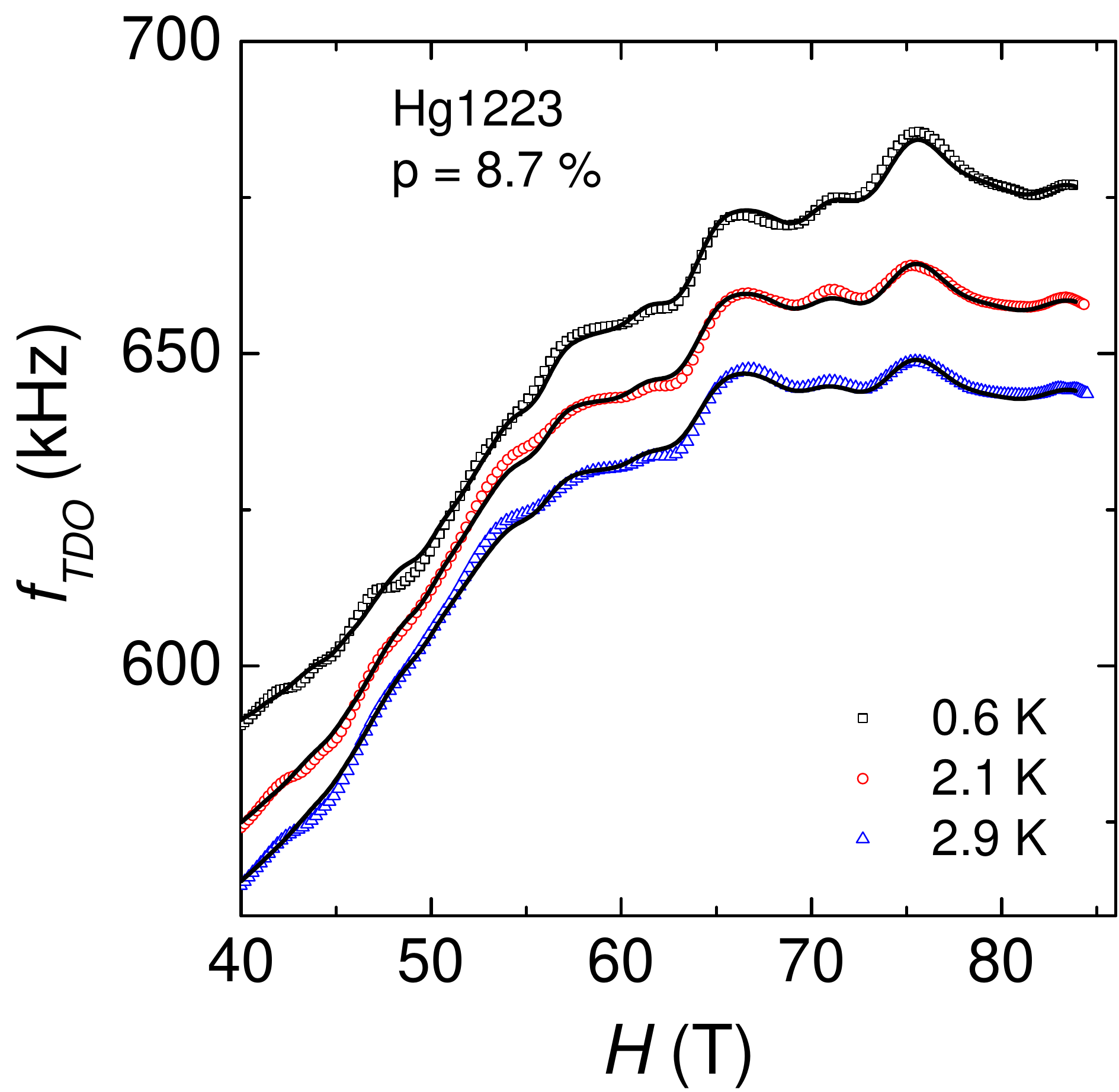}
\caption{\textbf{Lifshitz-Kosevich fit.} Field dependence of the TDO frequency in Hg1223 ($p$ = 8.7~\%) at different temperatures (symbols). Solid lines correspond to the fit to the data using the Lifshitz-Kosevich theory plus a polynomial background in the field range $40 \leq H \leq 85$~T (see text). 
}
\label{FigS3}
\end{figure}
%%%%%%%%%%%%%%%%%%%%%%%%%%%%%%%%%%%%%%%%%%%
%
The same procedure was carried out for the $p$ = 8.7~\% sample in the temperature range $T$=0.6~K to $T$=2.9~K and in the field range $H$ = 40 - 85~T (see Fig.~\ref{FigS3}). Table~\ref{tabS3} shows the relevant parameters deduced from the fit, i.e. frequencies, effective masses and Dingle temperatures. Again, the oscillation frequencies  match the values obtained by the DFT within the error bars (see table~\ref{tab1} of the main text).

%%%%%%%%%%%%%%%%%%%%%%   TABLE S3  %%%%%%%%%%%%%%%%%%%%%
\begin{table}[]
\begin{tabular}{c|c|c|c|c|c|c|}
\cline{2-7}
                       & \multicolumn{3}{c|}{$p$ = 8~\%} & \multicolumn{3}{c|}{$p$ = 8.7~\%} \\ \cline{2-7} 
                       & $F_1$  & $F_2$ & $F_3$ & $F_1$  & $F_2$ & $F_3$ \\ \hline
\multicolumn{1}{|c|}{$F$ (T)}  & 331 & 500 & 866 & 333 & 514 & 820      \\ \hline
\multicolumn{1}{|c|}{$m_{c}^*$ ($m_{e}$)}  & 1.9 & 1.2 & 1.6 & 1.6 & 2 & 2.2 \\ \hline
\multicolumn{1}{|c|}{$T_D$ (K)}  & 8 & 7 & 14 & 5 & 7 & 16 \\ \hline
\end{tabular}
\caption{\label{tabS3} QO frequencies, effective masses and Dingle temperatures deduced from the fit of the TDO signal to the Lifshitz-Kosevich theory (see text) for the $p$ = 8~\% sample (see Fig.~2) and the $p$ = 8.7~\% sample (see Fig.~\ref{FigS3}).}
\end{table}

\section{Hall effect data}
\begin{figure} [] 
\centering
\includegraphics[width=0.45\textwidth]{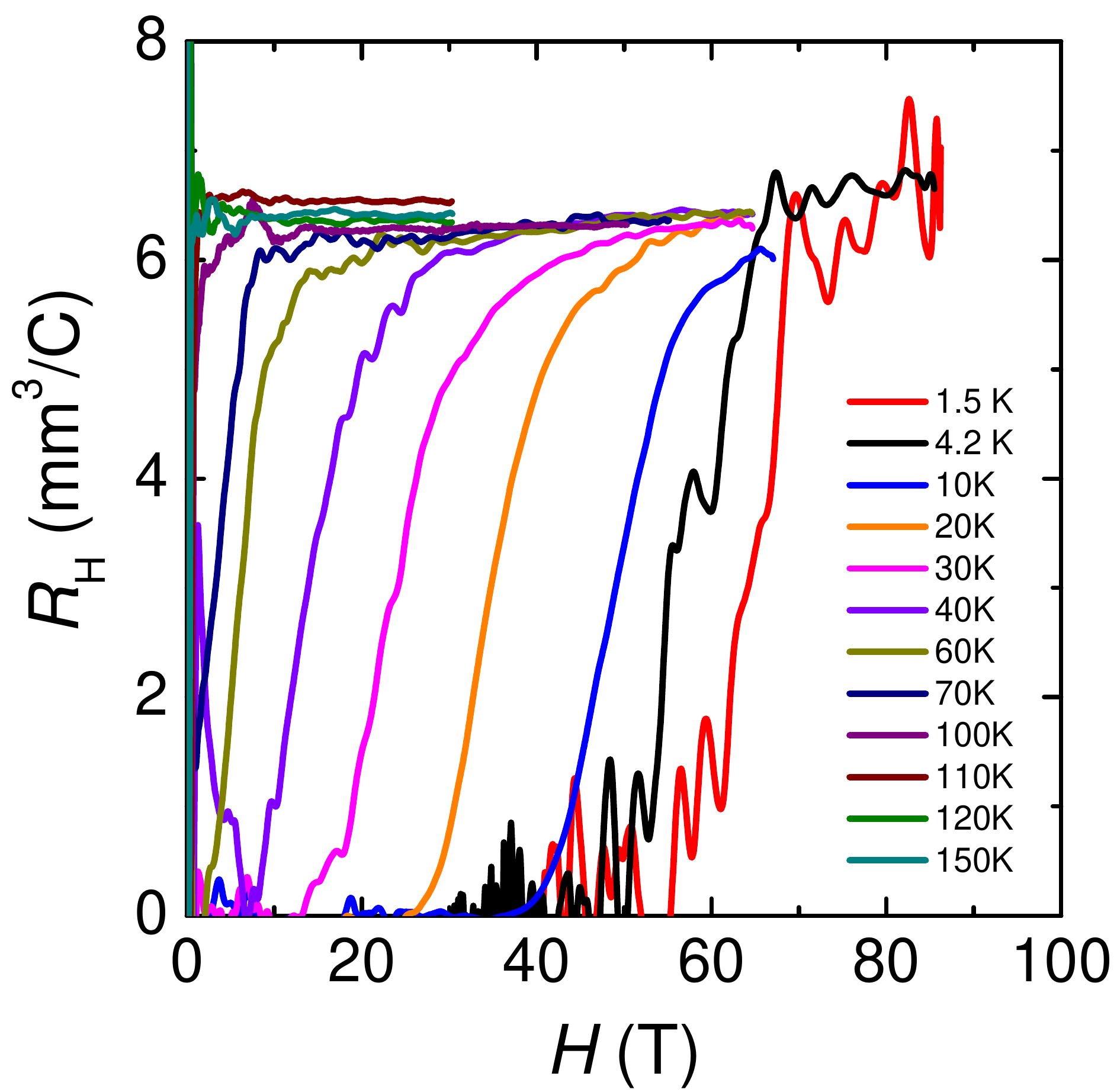}
\caption{\textbf{Field dependence of the Hall coefficient in Hg1223.}  Hall coefficient ($R_H$) of Hg1223 ($p$ = 8.8~\%) at various fixed temperatures, as indicated, versus magnetic field. 
}
\label{FigS4}
\end{figure}

The transverse Hall resistance $R_{xy}$ of our Hg1223 $p$ = 8.8~\% sample was measured in pulsed fields. 
Fig.~\ref{FigS4} shows the isotherms of the Hall coefficient, $R_H = t R_{xy} / H$, as indicated, where $t$ is the thickness of the sample. Data in the temperature range $T$ = 10 - 150~K was obtained in a 68~T pulsed magnet, where the signal to noise ratio was good. However, at $T$ = 10~K, 68~T was not enough to reach the normal state value of the Hall coefficient. We have thus performed additional measurements in a dual coil magnet to produce non-destructive magnetic fields up to 88~T at lower temperature, $T$ = 4.2~K and $T$ = 1.5~K. Due to larger d$H$/d$t$, the signal to noise ratio is worse but we can clearly state that $R_H$ remains positive down to the lowest temperature. This means that the Fermi surface of Hg1223 contains at least one mobile hole pocket at this doping level (see next section).

%%%%%%%%%%%%%%%%%%%%%%   FIGURE S4  %%%%%%%%%%%%%%%%%%%%%
%%%%%%%%%%%%%%%%%%%%%%%%%%%%%%%%%%%%%%%%%%%

\section{two-band model}
In the main text, we argue that one scenario compatible with the QO spectrum consists of antiferromagnetism (hole pocket) in the inner plane and charge order (electron pocket) in the outer plane (see Fig. 4b). Since the inner plane is protected from out-of-plane disorder, we assume that the hole-like carriers are more mobile than the electron-like carriers. 

%%%%%%%%%%%%%%%%%%%%%%   FIGURE S5  %%%%%%%%%%%%%%%%%%%%%
\begin{figure} [] 
\centering
\includegraphics[width=0.45\textwidth]{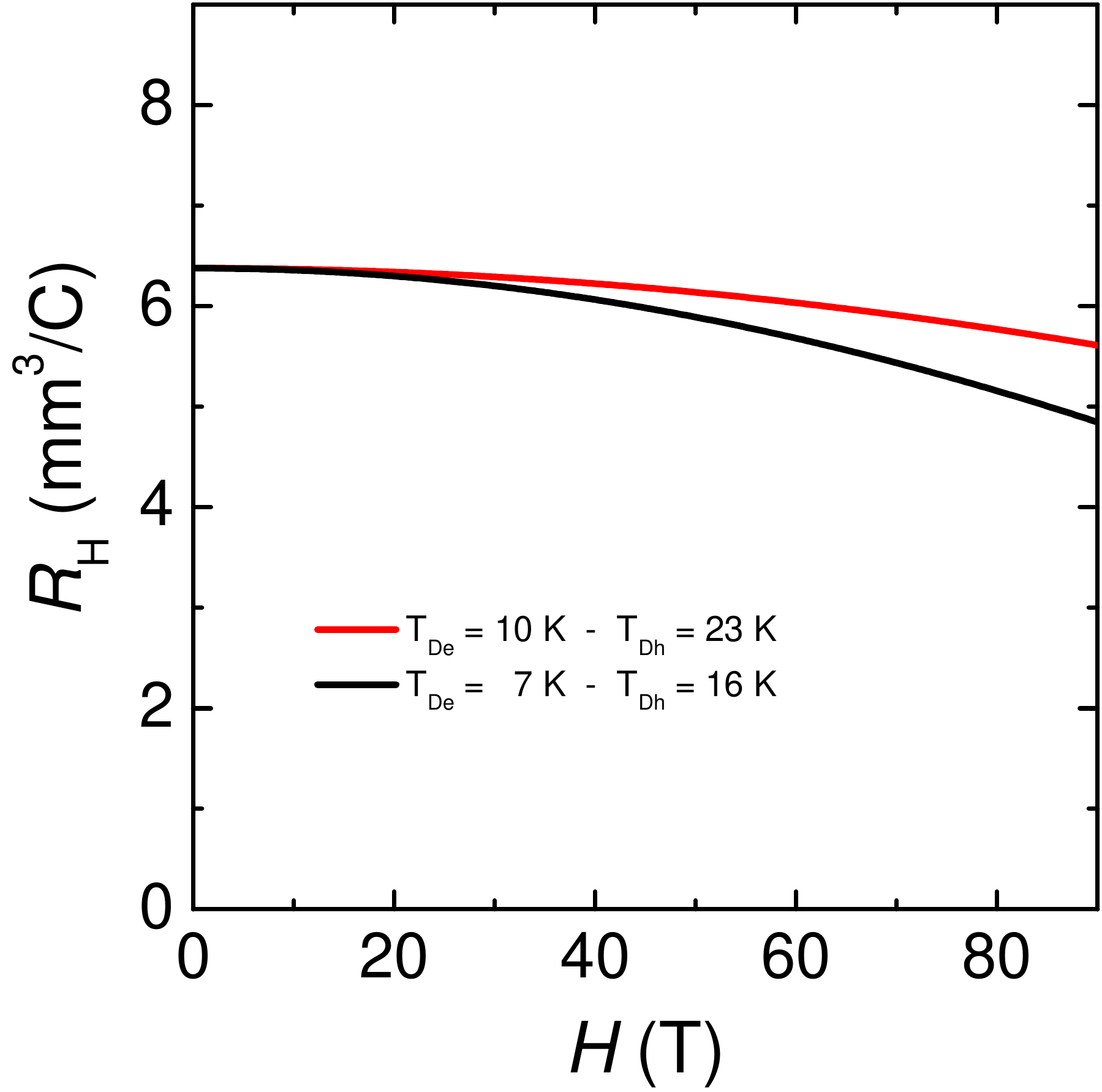}
\caption{\textbf{Field dependence of the Hall coefficient in the two-band model.}  Simulated $R_H$ versus magnetic field in the two-band model (see text). We use $m_{ce}$ = 2.5~$m_e$ and $m_{ch}$ = 2~$m_e$ for the effective mass of the electron (inner plane) and the hole (outer planes), respectively. The Dingle temperatures of the electron and the hole are indicated. 
}
\label{FigS5}
\end{figure}
%%%%%%%%%%%%%%%%%%%%%%%%%%%%%%%%%%%%%%%%%%%

Using a simple Drude model in the low field limit, the Hall coefficient is given by: 
\begin{equation}
R_H =\frac{\sigma_h \mu_h - \sigma_e \mu_e}{(\sigma_h + \sigma_e)^2}
\end{equation}
where $\sigma$ and $\mu$ are the conductivities and mobilities, respectively. 
To calculate the carrier density, we rely on the measured QO frequencies. 
According to Onsager's relation, the size of the Fermi surface pocket $A_\text{FS}$ in reciprocal space is :
	\begin{equation}\label{eq:}
		A_\text{FS}=\frac{2\pi e F}{\hbar},
	\end{equation} 
where $F$ is the SdH oscillation frequency.
The number of carriers per layer is given by:	
	\begin{equation}\label{eq:}
		N =\frac{2 A_\text{FS}}{A_\text{FBZ}},
	\end{equation} 
where $A_\text{FBZ}$ is the size of the first Brillouin zone in the reciprocal space.
In a unit cell there are three layers with one Cu atom per layer. We suppose that the Fermi surface of the outer planes contains a single electron pocket due Fermi surface reconstruction by charge order. The Fermi surface of the inner layer contains two hole pockets due to Fermi surface reconstruction by antiferromagnetism. Then the carrier density $n_{h(el)}$ of holes(electrons) is calculated according to the following formula:
	\begin{equation}%\label{eq:}
		n_\text{h} =\frac{N_\text{h}}{V}, 	n_\text{el} =\frac{2 N_\text{el}}{V} 
	\end{equation} 
where $V=236.9\text{ \AA}^3$ is the unit cell volume of Hg1223.
As stated in the main text, a Hall coefficient $R_H \approx$ 6.5 mm$^3$/C (see Fig.~3) corresponds to a ratio of mobilities $\mu_h / \mu_e = $2.85, a reasonable value owing to the disorder-protected nature of inner plane compare to outer plane. 
Now we have to justify the low-field limit. This is in fact indicated in the data of Fig.~\ref{FigS4} by the slight field dependence of $R_H$. In the field-dependent two-band model, $R_H$ is given by:
\begin{equation}\label{eq:r_hall}
		R_{H}(H)=
		\frac
		{\sigma_{h}^2 R_{h} + \sigma_{e}^2 R_{e}+
			\sigma_{h}^2 \sigma_{e}^2 R_{h} R_{e}
			(R_{h} + R_{e})H^2}
		{(\sigma_{h}+\sigma_{e})^2+
			\sigma_{h}^2\sigma_{e}^2(R_{h}+R_{e})^2H^2}
	\end{equation}
where $\sigma_{h(e)}$ and $R_{h(e)}$ are the conductivity and Hall number of the $h(e)$ carriers, respectively: 
$\sigma_{h(e)}=en_{h(e)}\mu_{h(e)}$ and $R_{h(e)}= \frac{1}{e n_{h(e)}}$.	

Fig.~\ref{FigS5} shows the simulated Hall coefficient using a set of parameters (effective masses and Dingle temperatures) compatible with the fitting procedure shown in Fig.~\ref{FigS3} for the $p$ = 8.8~\% sample. We can see a slight field dependence of $R_H$ that is not in disagreement with the experimental data given the signal to noise ratio at low temperature and the limited field range. Note that this model is very naive since it does not take into account the anisotropy of the scattering rate along the Fermi surface, that can be substantial in the cuprates.

\section{Sketch of Fermi surface}
%%%%%%%%%%%%%%%%%%%%%%   FIGURE S6  %%%%%%%%%%%%%%%%%%%%%

%%%%%%%%%%%%%%%%%%%%%%%%%%%%%%%%%%%%%%%%%%%
%

\begin{figure} [] 
\centering
\includegraphics[clip, trim=3cm 0.5cm 0.5cm 0.5cm, width=0.5\textwidth]{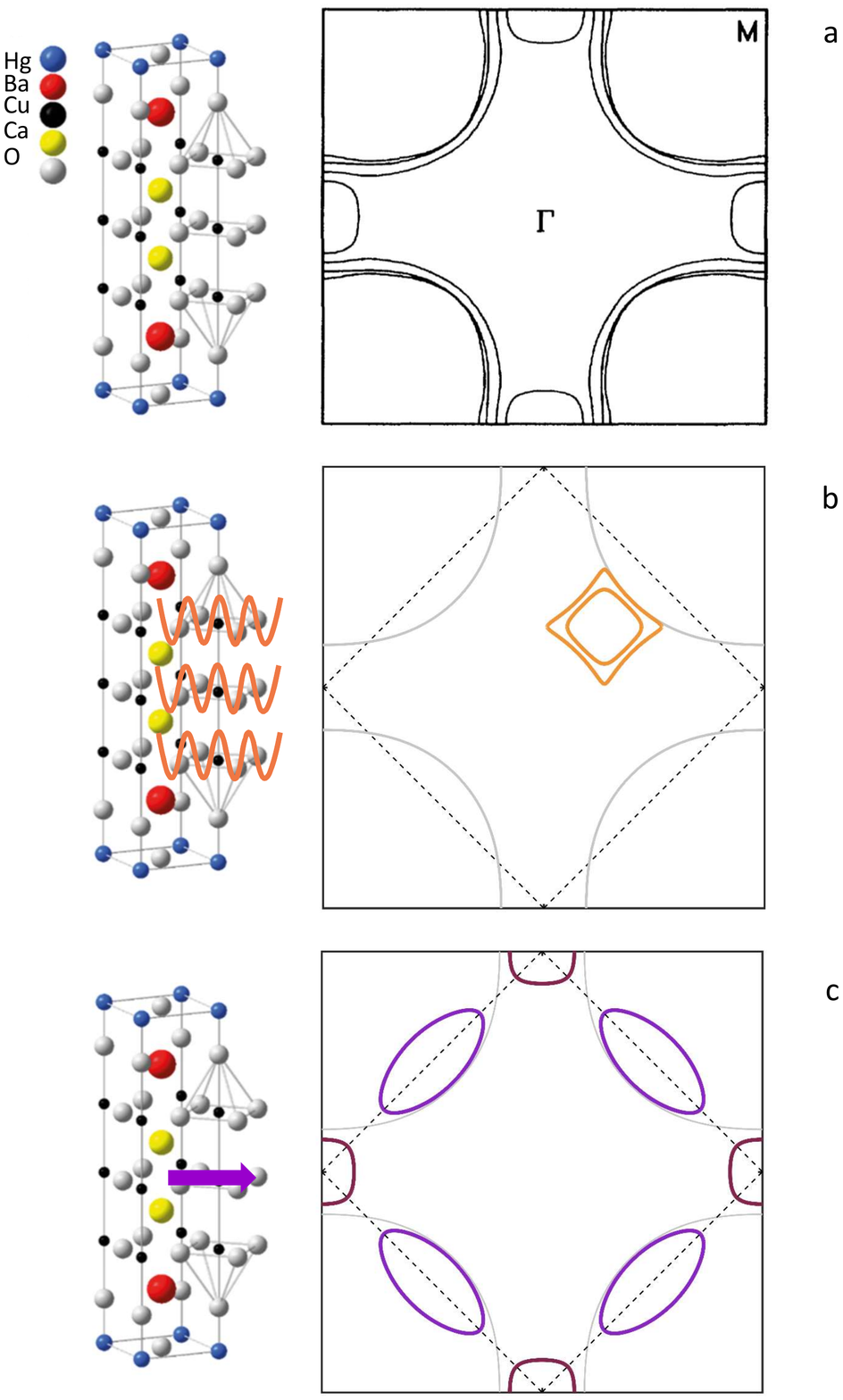}
\caption{\textbf{Sketch of the Fermi surface.}  To illustrate the discussion in the main regarding the different scenarios, we show a) FS obtained by LDA calculation (from Ref.~\onlinecite{Singh93}). b) Sketch of the FS assuming a Fermi surface reconstruction by charge order in the three CuO$_2$ planes. c) Fermi surface of the inner plane assuming a Fermi surface reconstruction by antiferromagnetism.   
}
\label{FigS6}
\end{figure}

The sketches presented in Fig.~4 and Fig.~\ref{FigS6} are based on calculations of Fermi surface reconstruction due to AFM order and/or CDW order. The starting point of the calculations is the unreconstructed Fermi surface of the single-layer Hg1201 compound, parametrized by the tight-binding equation: 

\begin{multline}\label{eq:Ek}  
		E(k)=-2t_{1}(cos(ak_{x})+cos(bk_{y}))\\ - 4t_{2}(cos(ak_{x}) \times cos(bk_{y})) - 2t_{3}(cos(2ak_{x}) + cos(2bk_{y}))\\ -4t_{4}(cos(2ak_{x}) \times cos(bk_{y}) + cos(ak_{x}) \times cos(2bk_{y})) - \mu
\end{multline}

where $(t_{1},t_{2},t_{3},t_{4})=(0.48, -0.105, 0.08, -0.02)$ eV, $k_{x}$ and $k_{y}$ are the in-plane wavevectors, $a$ and $b$ the in-plane lattice constants. The tight-binding parameters used in the calculations are chosen in order to reproduce the photoemission data reported in Hg1201 \cite{Vishik}. The chemical potential ($\mu$) is adjusted to yield a doping level p =~8.5\% (for $\mu$ = -0.355 eV) so that the total carrier density is equal to 1 + $p$ holes, according to the Luttinger sum rule. 

For the calculation of Fermi surface reconstruction due to AFM order, we used the following analytical formula for the electron ($E_{e}(k)$) and hole ($E_{h}(k)$) band structure:

\begin{multline}\label{eq:Ee}
		E_{e}(k)=
		\frac
		{E(k)+E(k+Q_{AF})}
		{2}\\ + \sqrt{\frac
		{((k)+E(k+Q_{AF}))^2}{4}
		+V_{AF}^2
		}
	\end{multline}
	
	\begin{multline}\label{eq:Eh}
		E_{h}(k)=
		\frac
		{E(k)+E(k+Q_{AF})}
		{2}\\ - \sqrt{\frac
		{((k)+E(k+Q_{AF}))^2}{4}
		+V_{AF}^2
		}
	\end{multline}

where $V_{AF}$ is the AFM potential, $Q_{AF} = (\pi,\pi)$ is the antiferromagnetic wavevector. The hole pocket (purple FS in Fig.~4) whose area corresponds to a frequency of 500~T is obtained using $V_{AF}$ = 0.445~eV. 

For the calculation of Fermi surface reconstruction due to bi-axial charge order, only the first-order translations of the band-structure due to charge-ordering with a characteristic wavevector $Q_{CDW}$ are considered \cite{Harrison}. The resulting $4\times 4$ Hamiltonian is then diagonalized numerically, with a typical mesh of $10^{6}$ points. We choose $Q_{CDW}$ = 0.275 r.l.u and $V_{CDW}$  = 0.075 eV to produce the electron pocket with an associated frequency of 850 T displayed as an orange line 
in Fig.~4.

%%%%%%%%%%%%%%%%%%%%%%%%%%%%%%%%%%%%%%%%%%%%%%%

%%%%%%%%%%%%%%%%%%%%%%%%%%%%%%%%%%%%%%%%%%%%%%%

Fig.~\ref{FigS6}a reproduces the Fermi surface of the stoichiometric compound HgBa$_2$Ca$_2$Cu$_3$O$_8$ obtained by local density approximation calculations \cite{Singh93}. It consists of three large hole-like tubular CuO$_2$ sheets centered on the corner of the Brillouin zone plus a small electron-like Fermi surface located at the anti-node. The latter disappears with doping \cite{Singh93}. The size of the large orbits translates into a QO frequency of about 15~kT, a value much larger than the observed frequencies.\\
Fig.~\ref{FigS6}b corresponds to the Fermi surface of Hg1223 assuming a Fermi surface reconstruction by charge order in the three CuO$_2$ planes. Here, we assume that the Fermi surfaces derived from the outer planes are similar in size. Therefore, a Fermi surface reconstruction by charge order leads to two electron pockets located at the node. For the sake of simplicity, solely the strength of the CDW potential $V_{CDW}$ was adjusted to produce the electron pockets with $F$ = 850~T and $F$ = 500~T displayed in Fig~\ref{FigS6}b. Similar results could be obtained by varying the CDW wavevector $Q_{CDW}$.  \\
Fig.~\ref{FigS6}c depicts the Fermi surface of the inner plane assuming a Fermi surface reconstruction by AFM order. 
The hole (electron) FS presented as purple (wine) lines in Fig.~\ref{FigS6}c corresponding to a quantum oscillation frequency of $F_h$ = 850~T ($F_e$ = 500T) are obtained using $V_{AF} = 0.25$ eV. They are compatible with the observed QO frequencies $F_2$ and $F_3$. The third observed frequency ($F_1$ = 350 T) could then be due to magnetic breakdown between these two orbits. However, the magnetic breakdown field $B_0$ required to enable magnetic breakdown orbits is $B_0 \propto \Delta k^2$ where $\Delta k$ is the separation between the two orbits.  This large gap precludes the observation of magnetic breakdown in the QOs. A simple estimate of the magnetic breakdown field can be achieved using the Blount criterion: 
%It consists of both hole ($F_h$ = 850 T, magenta line) and electron ($F_e$ = 500 T, bordeaux colour line) pockets whose size 

\begin{equation}\label{eq:Blount}
		B_{0}=\frac{m^{*}}{e\hbar} \times \frac{E_{g}^{2}}{E_{F}}
	\end{equation}

where $E_{F}=\frac{e\hbar}{m^{*}} \times F_{QO}$ is the Fermi energy, $E_g$ is the energy gap due to AFM order, $m^{*}$ the effective mass and  $F_{QO}$ the QO frequency. Using $E_g$ = 0.445 eV, $m^{*}$ = 1.6$m_{0}$ (1.2$m_{0}$) and $F_{QO}$ = 850 T (500 T) yields an estimated magnetic breakdown field $B_0 \approx$ 13000 T (14000T),  more than two orders of magnitude larger than the magnetic field used in the present study. 
%%%

\end{document}